\documentclass[prl,twocolumn,amsmath,amssymb]{revtex4}
\usepackage{graphicx}
\usepackage{amssymb,amsmath}
\begin{document}

\title{Massively parallel single-molecule manipulation using centrifugal force}

\author{Ken Halvorsen}
\author{Wesley P. Wong}
\email{wong@rowland.harvard.edu}
\affiliation{The Rowland Institute at Harvard, Harvard University, Cambridge, Massachusetts 02142, USA}

\date{\today}

\begin{abstract}

Precise manipulation of single molecules has already led to remarkable insights in physics, chemistry, biology and medicine. However, widespread adoption of single-molecule techniques has been impeded by equipment cost and the laborious nature of making measurements one molecule at a time. We have solved these issues with a new approach: massively parallel single-molecule force measurements using centrifugal force. This approach is realized in a novel instrument that we call the Centrifuge Force Microscope (CFM), in which objects in an orbiting sample are subjected to a calibration-free, macroscopically uniform force-field while their micro-to-nanoscopic motions are observed. We demonstrate high-throughput single-molecule force spectroscopy with this technique by performing thousands of rupture experiments in parallel, characterizing force-dependent unbinding kinetics of an antibody-antigen pair in minutes rather than days. Additionally, we verify the force accuracy of the instrument by measuring the well-established DNA overstretching transition at 66 $\pm$ 3 pN. With significant benefits in efficiency, cost, simplicity, and versatility, ``single-molecule centrifugation'' has the potential to revolutionize single-molecule experimentation, and open access to a wider range of researchers and experimental systems.

\end{abstract}

\maketitle

Single-molecule research has advanced greatly in the last decade, fuelled largely by the development of technologies such as the AFM and optical and magnetic tweezers, which enable precise physical manipulation of single molecular constructs \cite{neuman2008smf}. Remarkable studies with these instruments have already yielded new insight into such diverse areas as protein folding and unfolding dynamics \cite{kellermayer1997fut, rief1997rui}, motor proteins \cite{block1990bms,mehta1999mvp}, dynamic strength of receptor ligand interactions \cite{merkel1999elr, evans2001cdt}, enzymatic activity \cite{zhangvwf2009} and DNA mechanics \cite{bustamante2003tyt}.  Widespread use of these powerful techniques, however, has been impeded by the laborious nature of making measurements one molecule at a time, the typically costly equipment, and the requisite technical expertise to perform these measurements. Recently these issues have received some attention with innovations such as multiplexed magnetic tweezer systems \cite{danilowicz2005dissociation,ribeck2008multiplexed} to increase efficiency, and more cost-effective designs for optical tweezers systems \cite{smith1999inexpensive}.

\begin{figure}[tbp]
  \begin{center}
    \includegraphics[width=.45\textwidth]
    {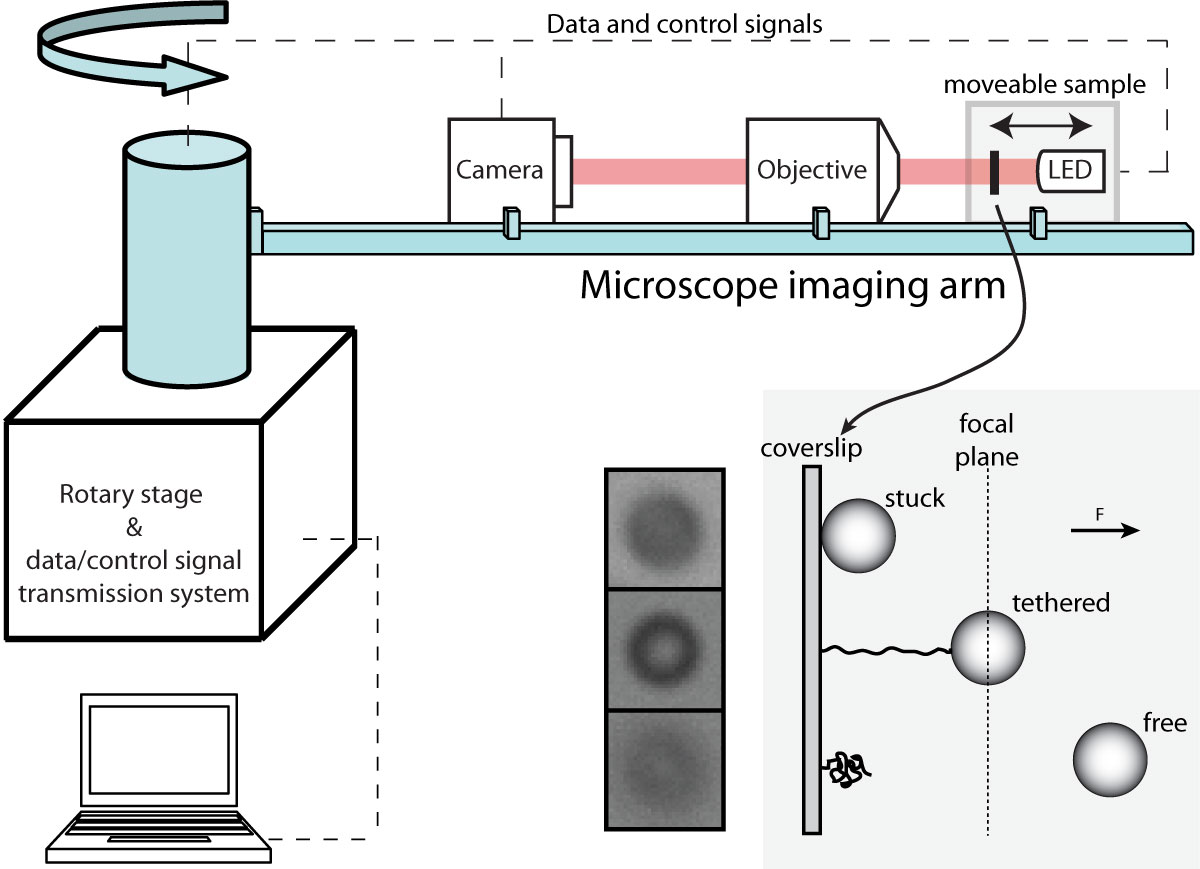}
    \caption{\label{fig:schematic}The centrifuge force microscope.  A rotary stage spins a miniaturized microscope, imparting a centrifugal force on beads interacting with a coverslip (\emph{inset, right}).  Transmitted light microscope images are sufficient to clearly distinguish between stuck, tethered, and untethered beads (\emph{inset, left}) using a 20x lens, 24kB DNA, and 2.8 micron beads.  Dynamic readout and control of the CCD, LED, and piezo translator during rotation is enabled by an integrated fiber optic rotary joint with electrical slipring (not shown).}
  \end{center}
\end{figure}

We have developed a new approach to solve these problems: massively parallel single-molecule force measurements using centrifugal force. The basic concept is that by rapidly rotating a high-resolution detection system, a centrifugal force field can be applied to an ensemble of objects while simultaneously observing their micro-to-nanoscopic motions. This is implemented in a new instrument that we call the Centrifuge Force Microscope (CFM) (Figure \ref{fig:schematic}), where an entire miniaturized video light microscope is mounted to a rotary stage. High-throughput single-molecule force spectroscopy is achieved by linking beads to a coverslip with single-molecule tethers and orienting the surface normal to the applied centrifugal force. This differs from previous centrifuge microscope instruments in which the centrifugal force is applied parallel to the coverslip/substrate \cite{oiwa1990steady,hall1993force,inoue2001centrifuge}. By pulling the tethered particles directly away from the substrate, lever arm effects are minimized and control over surface-surface interactions is increased (e.g. controlling the Òtouch timeÓ over which functionalized surfaces are in contact), enabling precise single-molecule force measurements.

\begin{figure*}[tb]
  \begin{center}
    \includegraphics[width=0.75\textwidth]
    {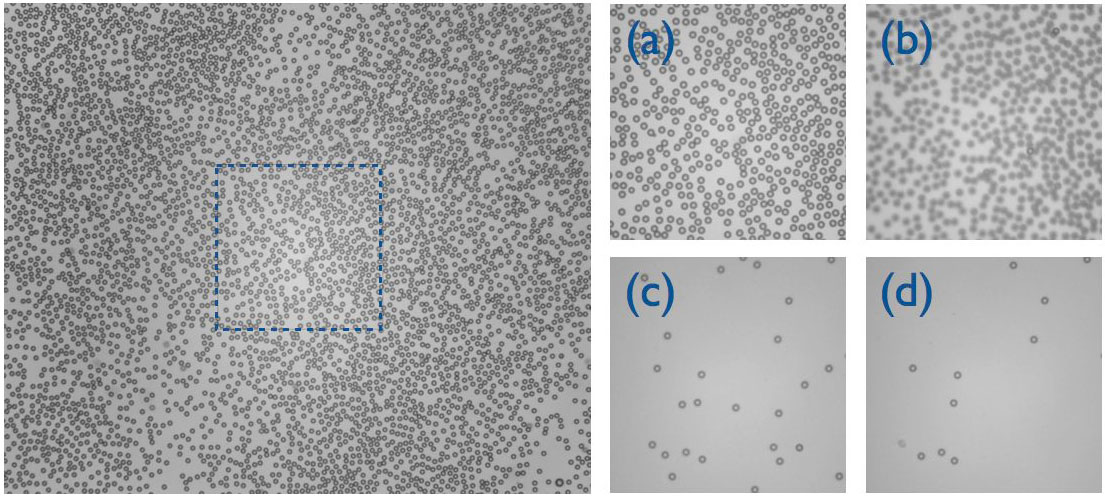}
    \caption{\label{fig:screenshots}Images from a CFM experiment.  Time lapse images show the progression of a bond rupture experiment. Thousands of receptor-functionalized beads (diameter 2.8 microns) can be seen near the ligand-functionalized coverslip (\emph{left}). Zoom-in of a smaller region at four different times (\emph{right}). (a) Beginning of the experiment with beads resting against the coverslip (b) The objective is focused one-tether-length away from the coverslip into the sample, so all beads in the wrong focal plane are blurry. (c) A centrifugal force field is generated that pulls the beads away from the coverslip. Beads tethered by a single DNA molecule move into focus, while unattached beads leave the field of view. (d) Beads detach from the coverslip as single receptor-ligand bonds rupture, resulting in fewer visible beads over time.}
  \end{center}
\end{figure*}

The centrifugal force applied to each molecular tether can be easily determined using $F = m \omega^2 R$, where $m$ is the mass of the bead (minus the mass of the medium displaced to account for buoyancy), $\omega$ is the magnitude of its angular velocity, and $R$ is its distance from the axis of rotation. Since $R$ is a macroscopic length much larger than the motion of the particles and the region of observation, the force field is conveniently uniform over the sample and as stable as the constancy of $\omega$. For monodisperse beads of known size and density, the centrifugal force on each particle is identical and can be calculated directly without calibration. Detection of molecular transitions, such as bond rupture or tether extension, is also straightforward. The coverslip is oriented so that tethered beads are pulled away from the camera. Since the whole system rotates, the beads appear relatively stationary in the field of view, but are pulled out of focus as a molecular tether stretches or detaches.  While a variety of bead detection schemes are possible, image focus provides the simplest way to determine if a bead is connected to the surface or not.  For example, when measuring bond dissociation kinetics under constant force, one simply needs to measure the times at which singly tethered beads abruptly detach from the coverslip and disappear from view.

We demonstrate this method by performing thousands of single-molecule measurements in parallel to characterize the force-dependent unbinding kinetics of an antibody-antigen interaction (monoclonal anti-digoxigenin with digoxigenin). This is illustrated in Figure \ref{fig:screenshots}. Antigen molecules were tethered to monodisperse 2.8 micron beads by DNA tethers and brought into contact with the anti-digoxigenin coated coverslip. The sample was then accelerated within a few seconds to a constant velocity in order to apply a uniform force field to all of the beads. Singly-tethered beads responded by moving away from the glass substrate by a distance consistent with the compliance of double-stranded DNA, whereas beads that were not tethered to the surface extended far out of focus, and non-specifically bound beads remained at or near the glass surface. Thus, the DNA tether provided a molecular signature for the discrimination of single molecular tethers, which could easily be observed by bead focus. Bond rupture events, indicated by detachment of tethered beads, were visually dramatic, and could be distinguished automatically by standard image processing algorithms.

\begin{figure}[tb]
  \begin{center}
    \includegraphics[width=.45\textwidth]
    {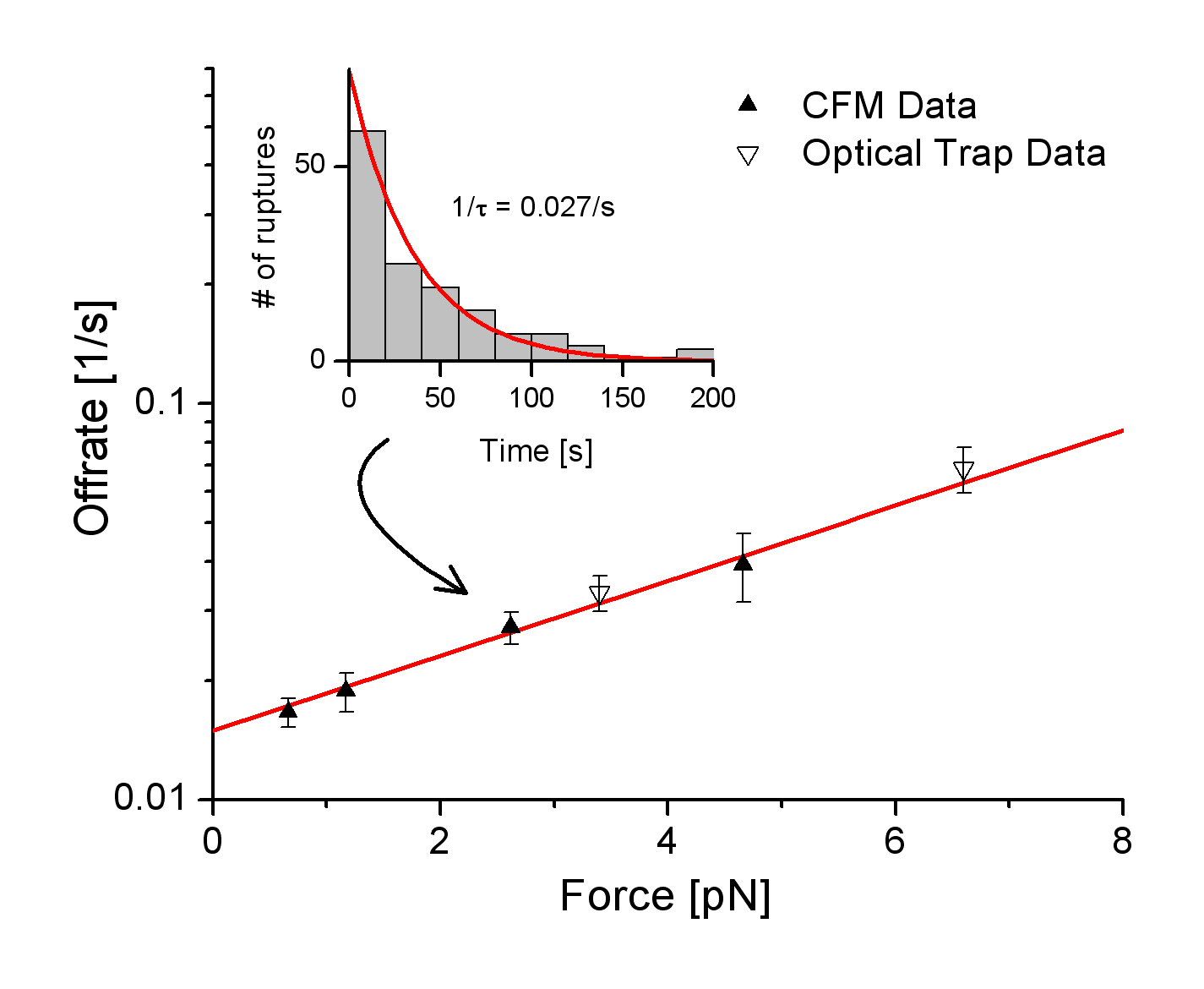}
    \caption{\label{fig:dig-anti-dig} Force-dependent unbinding of digoxigenin and its antibody.  Force clamps ranging from hundreds of femtoNewtons to several picoNewtons were applied using the CFM (filled triangles), as well as with the optical trap (open triangles).  Each CFM data point was obtained from a single experiment lasting a few minutes, while optical trap data was collected serially over a period of many hours.  Histograms of the rupture times with a 20 second bin width (10 seconds for the highest optical trap force) were fit with a decaying exponential to obtain the off-rate at each force (\emph{inset}). Plotted error bars in the off-rate result from the uncertainty in the least squares fit.}
  \end{center}
\end{figure}

By applying various force clamps, we determined the force-dependent off-rate $k_\text{off} (f) = k_0 \exp(f/f_\beta)$ \cite{bell1978models,evans1997dynamic} for the interaction of digoxigenin and its antibody. We found a stress-free off-rate of $k_0$ = 0.015 $\pm$ 0.002 s$^{-1}$ and a force scale of $f_\beta$ = 4.6 $\pm$ 1.3 pN  (Figure \ref{fig:dig-anti-dig}). Using the same construct, we applied force clamps using our micropipette-based optical trap force probe (instrument and methodology described previously in Reference \cite{zhangvwf2009}) and recorded rupture times, finding near perfect agreement with CFM measurements. Additionally, these results agree within error with previous AFM experiments \cite{neuert2006dynamic}. As an additional verification of the instrument, we used 25 micron beads to overstretch DNA, and found that overstretching occurred at 66 pN $\pm$ 3 pN, in agreement with previous measurements \cite{smith1996overstretching,cluzel1996dna}.

As demonstrated by these examples, the CFM offers a unique set of advantages, largely derived from the properties of the centrifugal force field, namely that it is macroscopically uniform, highly stable, calibration-free, and dynamically controllable in an essentially deterministic way. Thus, a desired force history can be applied to an ensemble of single molecules without the need for active feedback. The force field conveniently couples to mass density, complimenting optical and magnetic tweezers that couple to polarizability and magnetic moment. Not only does this eliminate the possibility of radiative damage, but it also expands the range of systems that can be studied with force (e.g. beads or objects made of any material can be used, as long as they have a different mass density than their surroundings). Furthermore, the CFM can achieve an enormous force range by varying bead size, bead material, and rotation speed. For example, even our current prototype is capable of sub-femtoNewton (e.g. 1 micron polystyrene particles in water, at 60 rpm) to microNewton (e.g. 100 micron borosilicate glass particles in water, at 600 rpm) forces, covering a range that would typically require more than one instrument \cite{neuman2008smf}.

The most obvious benefit of this method is the ability to perform massively parallel measurements, enabling dramatic improvements in efficiency capable of reducing experimental time from days to minutes.  More than simply speeding up progress, this efficiency also enables new experiments that would be all but impossible with other methods (e.g. near equilibrium measurements that observe interactions with hour-long lifetimes would by unfeasible with sequentially collected statistics). With larger statistical sets now easily attainable, more detailed characterizations and model testing is possible, as well as the observation of population heterogeneity.  Additionally, parallel measurements can be used to test entire families of interactions simultaneously (e.g. multiple drugs candidates could be tested simultaneously against a target receptor).
 
The CFM is also extremely versatile, able to incorporate many existing microscopy technique to suit individual needs.  For instance, reflection interference contrast microscopy (RICM) imaging \cite{heinrich2008ibi} could be incorporated to achieve sub-nm resolution bead tracking, and fluorescence imaging could be added to allow visualization of subtle molecular transitions during the experiment \cite{tarsa2007dfi, ha2001smf}.

Importantly, this method is more cost effective and simpler to use than many other common methods. The single-molecule centrifugation experiments are straightforward, with a pre-programmed force protocol, minimal setup, and no need for user intervention. The material cost of our prototype was $\sim$\$15,000, and could potentially be reduced to $\sim$\$5,000 by using a more cost effective rotary drive and camera, as these two items account for ~75\% of the cost (e.g. the high precision rotary positioning stage could be replaced with a simpler electric motor, and a lower resolution camera could be used).  Material cost for established methods capable of single molecule measurements such as the optical trap and AFM can easily escalate above \$100,000, though systems with more limited functionality can be significantly cheaper. 

Centrifugal force has been overlooked in the single-molecule community for too long. We have shown that``single-molecule centrifugation'' can be simple, inexpensive, and significantly more efficient than many methods, enabling new avenues of research and opening single-molecule experimentation to a wider scientific community.

\section{Materials and Methods}
\subsection{Instrument Description}
The Centrifuge Force Microscope consists of an optical system mounted perpendicularly to the rotation axis of a computer controlled rotary stage (ADRT-150, Aerotech and CP20, Soloist).  The optical system consists of an LED lamp (LED528E, Thorlabs), 20x plan objective (NA 0.4), and a 5 megapixel CCD camera (GC 2450, Prosilica; maximum frame rate 15 fps).  The sample holder is an acrylic chamber sealed with a coverslip and an o-ring.  It can be translated (i.e. focused) using an adjustment screw (AJS100-02H, Newport) for coarse adjustment and a piezo for fine adjustment (AE0505D08F, Thorlabs).  Data and power transmission between the rotating optical system and external devices is accomplished with a fiber optic rotary joint (MJX-155-28-SC, Princetel) integrated into an electrical slipring (SRF24, Princetel).  Two media converters (855-10734 and 855-10735, IMC Networks) convert the gigabit ethernet signal from the camera to fiber optic and back to gigabit ethernet for connection to a computer (Precision T5400, Dell).  Power for the camera, media converter, LED, and piezo are transmitted through the electrical slipring, allowing dynamic control while spinning. A counterweight placed opposite to the optical system balances the load. With a top speed of 600 rpm and a rotation arm length of 385.5 mm, the current instrument can generate a force field of $\sim$1--155 g.

\subsection{Experimental Details}
Digoxigenin labeled DNA was prepared by labeling the cohesive ends of 48kB lambda phage DNA (N3013S, New England Biolabs) with biotin (biotin-14-dATP and biotin-14-dCTP, Invitrogen) using Klenow polymerase (M0212S, New England Biolabs), followed by purification (QiaQuick purification kit, Qiagen).  The biotinylated DNA was then cut almost exactly in half with the XbaI restriction enzyme (R014SS, New England Biolabs) and re-purified.  The overhanging ends of the 24kB DNA were filled in and labeled with a single digoxigenin (dig-11-dUTP, Roche) or plain nucleotides for dig-labeled or unlabeled DNA.  Finally, the DNA was re-purified and the dig-DNA was mixed with unlabeled DNA in a 1:4 ratio before reacting with highly monodisperse streptavidin labeled beads (Dynabeads M-270, Invitrogen, measured CV = 1.7\%) for use in the experiment.

Functionalized coverslips were prepared by base washing a glass coverslip (immersed for 5 minutes in boiling solution of 1 part 30\% H$_2$O$_2$, 4 parts NH$_4$OH, and 19 parts d-H$_2$O), followed by adsorption of anti-digoxigenin (11094400, mouse monoclonal, Roche).  The coverslip was then washed with blocking buffer comprised of PBS with 0.1\% Tween 20 and 1mg/ml dephosphorylated alpha-casein (C8032, Sigma).  Experiments were performed in the same blocking buffer to decrease non-specific adhesion to the surface.

For DNA overstretching experiments, streptavidin was adsorbed onto 25 micron polystyrene beads (Thermo Scientific NIST traceable size standards 4225A, 24.61$\mu$m $\pm$ 0.22 $\mu$m mean diameter, CV = 1.1\%), then biotinylated lambda DNA was added. Streptavidin was adsorbed onto a glass coverslip, and experiments were performed in a buffer of 150 mM NaCl, 10 mM Tris, 1 mM EDTA. Density tables were used to find the buffer density of 1.0043 g/cm$^3$ \cite{lide2004crc}. The bead density was measured to be 1.0481 g/cm$^3$ by centrifuging beads suspended in fluids of known density and examining which way they settled (beads floated in 6.98\% NaCl solution but sunk in 6.95\% NaCl solution).

To observe the overstretching transition, the angular velocity of the stage was increased in stepwise increments of 20 degrees per second while the variance of each bead image was measured. Overstretching was easily detected by a dramatic change in the variance (i.e. focus) of the beads. The beginning of this transition occurred at 213 rpm, which corresponds to 66 pN.
%What was the step size?

\subsection{Data Analysis}
Analysis of videos recorded at 10 frames per second was performed by identifying the locations of fully tethered beads at a frame near the beginning of the movie (once full speed was reached) and analyzing small regions of interest at each location in subsequent frames to determine the time of bond rupture.  Finding the bead locations was done in ImageJ by first performing a background subtraction, then making a binary image and using the analyze particles tool to find the center positions and the variance of a region of interest around each bead.  The image variance is important because stuck and tethered beads can easily be distinguished from the variance, and stuck beads were excluded from analysis in this way.  This information is then fed to a custom Matlab program, which analyzes subsequent frames of the video to find the mean, variance, and centroid in the small regions of interest.  The rupture time for each bead was identified by a dramatic drop in the measured variance to near zero, corresponding to a grey, bead-free image. In rare cases where multiple drops in the variance were observed, the bead was excluded from analysis, as this typically indicated a multiple tether.

Once rupture times were collected, histograms were made and this data was fit with a decaying exponential with no offset.  The resulting time constants and associated fitting errors were used to determine the off-rate at each force level using a simple Arrhenius approach with force-dependence modeled in the standard way by a single, sharp barrier.

The force imparted on the beads due to centrifugation was calculated as $F = m \omega^2 R$, where $m$ is the mass of the bead (minus the mass of the medium displaced), $\omega$ is the magnitude of its angular velocity, and $R$ is its distance from the axis of rotation.  In our case, $R = 385.5$ mm, and $m = 6.9 \times 10^{-12}$ g (calculated using manufacturer's bead specifications of 2.8 micron diameter and 1.6 g/cm$^3$ density).  A small additional force from the bead's weight was included using the pythagorean theorem. The weight of the bead acts perpendicularly to the centrifugal force, and was added in quadrature when calculating the net force on the tether.

%\bibliographystyle{unsrt}
%\bibliography{biblio}

\begin{acknowledgments} This work was supported by the Rowland Junior Fellows program.  We thank Diane Schaak for help with sample preparation, discussions, and comments on the manuscript.  We thank C. Stokes for help integrating and troubleshooting electronics, and D. Rogers for help with the fabrication of the instrument.
\end{acknowledgments}

\end{document}